\documentclass[cameraready]{Interspeech}

\title{MedASR: An Open-Source Model for High-Accuracy Medical Dictation}

\author{Ke}{Wu}
\author{Ehsan}{Variani}
\author{Tom}{Bagby}
\author{Shashir}{Reddy}
\author{Rory}{Pilgrim}

\address{Google Inc}

\email{\{wuke, variani, tombagby, shashir, roryp\}@google.com}


\usepackage{comment}
\usepackage{stfloats}
\usepackage{amsmath}
\usepackage{amsfonts}
\usetikzlibrary{arrows.meta, decorations.pathreplacing, calc, shapes.geometric}

\begin{document}

\maketitle

\begin{abstract}
    We present MedASR, an open-source 105M-parameter model engineered for high-accuracy medical dictation. Prioritizing a ``small, fast, and accurate'' design, MedASR addresses 3 core pillars (1) Data: overcoming clinical corpora scarcity and class imbalance; (2) Modeling: efficient long-form training; and (3) Inference: accurate transcription via a pseudo-streaming sliding-window approach. Our evaluation shows that MedASR achieves a 58\% relative WER reduction on Eye Gaze compared to Whisper Large-v3. By open-sourcing MedASR, we provide a transparent, high-performance backbone for specialized healthcare applications, breaking down the barriers to clinical documentation often obscured by proprietary systems.
    \footnote{Submitted to \emph{Interspeech}.}
\end{abstract}

\section{Introduction}

The administrative burden of clinical documentation is a primary driver of physician burnout, creating an urgent need for robust Automated Speech Recognition (ASR) systems \cite{shanafelt2016relationship, arndt2017tethered}. While general-purpose foundation models have achieved remarkable versatility \cite{radford2023robust, team2023gemini}, they often lack the domain-specific grounding and structural awareness required for high-stakes medical reports. In this work, we introduce MedASR\footnote{https://huggingface.co/google/MedASR}, a 105M-parameter Conformer-based model \cite{gulati2020conformer} designed to be a high-performance starting point for medical voice technologies.

Our development was guided by an explicit goal that \emph{the model must be accurate, fast, and small}. Unlike ``monolithic'' foundation models that require massive cloud infrastructure, MedASR captures the complexities of medical nomenclature while remaining efficient for on-device deployment---a critical factor for preserving patient privacy by keeping sensitive clinical audio within the local infrastructure.

Historically, medical ASR has been dominated by proprietary, closed-source systems. This ``black box'' ecosystem restricts researchers from auditing model biases, refining clinical safety, or adapting models to niche sub-specialties. We believe that solving the fundamental problems of medical ASR requires collective community effort. By open-sourcing MedASR, we provide a transparent foundation for the next generation of clinical AI, moving away from vendor-locked solutions toward a collaborative, research-first framework.

The transition from general ASR to a medical system involves overcoming significant hurdles: 

\begin{description}
    \item[Data Challenges] (Scarcity and Acoustic Imbalance) High-quality medical audio is scarce due to stringent privacy and de-identification requirements. Furthermore, existing datasets often fail to balance ``general-domain'' acoustic fluency with specialized ``medical-domain'' nomenclature and physician-specific speaking styles. 

    \item[Modeling Challenges] (Long-Form Sequence Acceleration) Clinical dictations frequently exceed the typical 30-second window utilized by many general-purpose models, such as Whisper \cite{radford2023robust}. Training on such extended sequences on hardware accelerators introduces severe memory and batch-size constraints. The quadratic complexity of standard attention mechanisms makes it computationally difficult to scale to these lengths without specialized treatments, presenting a significant bottleneck for training ASR models on high-fidelity medical corpora.

    \item[Inference Challenges] (Long-Form Stability) At the inference stage, the requirement for precision is absolute. In a medical context, transcribing ``hypo-'' for ``hyper-'' is a safety failure. General-purpose models are prone to ``hallucinating'' clinically incorrect terms or experiencing ``drift'' during extended recordings---a known stability issue where the model fails to maintain alignment or begins deleting sequential content \cite{bain2023whisperx}. 
\end{description}

The following sections detail our architectural choices and the specific methodologies utilized to address each of these fundamental problems. We demonstrate that MedASR achieves superior performance over general-purpose foundational models like Whisper and Gemini, while further enabling alternative serving paradigms for on-device deployment and pseudo-streaming inference.

\section{The MedASR Foundation}

MedASR is built on a 105M-parameter Conformer architecture \cite{gulati2020conformer} and trained using 
a JAX-based \cite{jax2018github} framework. To address the development hurdles outlined previously, we implemented the following strategies:

\subsection{Data Scarcity, Acoustic Imbalance, and Formatting}
\label{sec:data}

The primary bottleneck in medical ASR is the acquisition of large-scale, high-fidelity audio that is both clinically relevant and acoustically diverse. We identified the following hurdles in data curation for a medical dictation model:

\begin{description}
    \item[Training Strategy and Specialty Coverage] To bridge the gap between general linguistic fluency and specialized medical expertise, we employ a two-step training pipeline: large-scale pre-training on general audio data followed by domain-specific fine-tuning. The proprietary dataset used for fine-tuning consists of 4,500+ hours of de-identified medical audio recordings and their corresponding transcriptions, primarily focused on physician dictations. This corpus covers four key medical specialties: \emph{Radiology} (RAD), \emph{Family Medicine} (FM), \emph{Internal Medicine} (IM), and \emph{General and Internal Medicine} (GENINT). The distribution across these areas ensures representation across diverse clinical vocabularies and reporting styles. Currently, MedASR is optimized for English-only environments, with plans for multilingual expansion in future iterations.
    \item[Choice of Pre-Training Data and Tokenization] Medical dictation requires ``print-ready'' formatting, including proper casing, punctuation, and mixed-format numbering. Most publicly available ASR datasets, such as LibriSpeech \cite{panayotov2015librispeech}, are heavily normalized. We chose LibriHeavy \cite{kang2023libriheavy} (non-normalized) for pre-training because it preserves these critical formatting features. For tokenization, we trained a SentencePiece \cite{kudo2018sentencepiece} model with a compact 512-vocabulary size, using a mixture of LibriHeavy and proprietary medical text. This small vocabulary was a deliberate choice to keep the model
    light and efficient for on-device serving. 
    \item[Analysis of Clinical Corpus] The statistics of our fine-tuning corpus are detailed in Table~\ref{tab:corpus_stats}. The data highlights a significant ``long-form'' challenge: while RAD dictations are relatively concise, specialties like GENINT and IM frequently exceed 1,000 seconds in duration. At the 99th percentile, a model must process sequences exceeding 6,000 tokens. 
\end{description}

\begin{table*}[b]
\caption{Statistics of the proprietary medical corpus. Percentiles (P90, P95, P99) for duration and token counts (512-vocab) illustrate the long-form challenge across specialties.}
\label{tab:corpus_stats}
\centering
\begin{tabular}{l rrr c rrr c rrr}
\toprule
& \multicolumn{3}{c}{Data Scale} && \multicolumn{3}{c}{Duration (seconds)} && \multicolumn{3}{c}{Token Count (512-Vocab)} \\
\cmidrule{2-4} \cmidrule{6-8} \cmidrule{10-12}
Specialty & Unique Spk & Avg Hr/Spk & Total Hr && P90 & P95 & P99 && P90 & P95 & P99 \\
\midrule
RAD    & 573 & 1.65 & 943.7 && 148.3 & 202.8 & 373.7 && 894 & 1182 & 2042 \\
FM     & 831 & 1.36 & 1130.6 && 530.0 & 674.0 & 1037.1 && 2928 & 3658 & 5151 \\
IM     & 1185 & 1.18 & 1402.0 && 578.0 & 724.8 & 1092.3 && 3244 & 3827 & 5328 \\
GENINT & 1265 & 0.86 & 1085.9 && 604.1 & 762.3 & 1209.5 && 3457 & 4132 & 6288 \\
\midrule
\textbf{Overall} & \textbf{3844} & \textbf{1.19} & \textbf{4562.1} && \textbf{441.4} & \textbf{575.1} & \textbf{933.0} && \textbf{2540} & \textbf{3221} & \textbf{4762} \\
\bottomrule
\end{tabular}
\end{table*}

\subsection{Modeling and Training Paradigm}

MedASR models the posterior probability of a token sequence $y$ given an input audio sequence $x$, denoted as $P_{\theta}(y|x)$. This section details our choices regarding modeling paradigms, network architecture, and optimization strategies.

\begin{description}
    \item[Posterior Model Selection] Several paradigms exist for modeling $P_{\theta}(y|x)$, including Cross-Entropy (CE) \cite{variani2017end}, Connectionist Temporal Classification (CTC) \cite{graves2012connectionist}, RNN-Transducer (RNN-T) \cite{graves2012sequence}, Hybrid Autoregressive Transducer (HAT) \cite{variani2020hybrid}, and Listen, Attend and Spell (LAS) \cite{chan2015listen}. As noted in \cite{variani2022global}, these models are fundamentally similar: each generates a posterior probability by marginalizing over alignment probabilities derived via the chain rule. Their primary distinctions lie in the probabilistic assumptions used to derive frame-level posteriors and the specific lattice structures used for alignment. For instance, CE and CTC objectives assume conditional independence between frames, enabling an encoder-only architecture. Conversely, models like RNN-T and LAS introduce dependencies on the prefix label history when modeling the probability of the next label, typically necessitating an encoder-decoder architecture. To maximize parallelization during both training and inference, we opted for an encoder-only architecture. We chose the CTC objective over the CE path to avoid the requirement for pre-aligned data, facilitating a more streamlined end-to-end training pipeline. In this framework, the encoder produces fixed-dimensional embeddings in $\mathbb{R}^d$, which are projected into a 512-token space. The model is optimized by marginalizing over all valid alignments in the CTC lattice $\mathcal{A}_{CTC}(x, y)$:
        \[
        L_{CTC} = -\log \sum_{z \in \mathcal{A}_{CTC}(x, y)} P_{\theta}(z | x)
        \]
        
    \item[Network Architecture] MedASR utilizes a 105M-parameter Conformer-L \cite{gulati2020conformer} backbone. Input audio is represented by 128-dimensional log-mel filterbank features extracted every 10ms with a 25ms window.
        \begin{itemize}
            \item Subsampling: Two 1D convolutions (stride 2, window size 5) reduce the encoder frame rate to 25Hz.
            \item Conformer Encoder: 17 layers, 512 activations, and 8 attention heads. Refinements: We deviate from the original implementation by using Rotary Positional Embeddings (RoPE) \cite{su2024roformer}, and removing biases in all layer normalization and dense layers to improve stability \cite{chowdhery2023palm}.
        \end{itemize}

    \item[Consistency Regularization] To enhance robustness, we employ Consistency Regularization CTC \cite{yao2024cr}. The encoder is run on two augmented versions of the input, $\tilde{x}_1$ and $\tilde{x}_2$, produced from independent applications of SpecAugment \cite{specaug2019} to the input $x$. The total loss is a weighted combination of the standard CTC loss averaged over both augmentations:
        \[
        L_{CTC} = \frac{1}{2}(L_{CTC}(\tilde{x}_1,y) + L_{CTC}(\tilde{x}_2,y))
        \]
        and a symmetric Kullback-Leibler (KL) divergence used as a regularization term:
        \[
        L_{reg} = D_{KL}(P_{\theta}(y|\tilde{x}_1) || P_{\theta}(y|\tilde{x}_2)) + D_{KL}(P_{\theta}(y|\tilde{x}_2) || P_{\theta}(y|\tilde{x}_1))
        \]
    We set the regularization weight to 0.2, and losses are averaged within the batch on a per-sequence basis.

    \item[Iterative Segmentation and Training] To accommodate the memory and computational constraints of hardware accelerators, fine-tuning sequences were limited to a maximum duration of 20 seconds (500 encoder frames). Given that clinical dictations often significantly exceed this limit, we developed a multi-stage iterative segmentation process to generate high-quality training pairs from long-form audio:
        \begin{enumerate}
            \item Bootstrapping: A seed model was trained exclusively on a subset of the data containing naturally occurring short sequences (up to 36s).
            \item Forced Alignment: This seed model was utilized to perform forced alignment on a fused CTC lattice (Section \ref{sec:pseudo-streaming}) from sliding windows of length 20s \& stride 18s, over the entire 4500-hour fine-tuning corpus.
            \item Lattice-Based Segmentation: At every 500 encoder frame mark, we extract the corresponding audio chunk and aligned text as a segmented example.
        \end{enumerate}
    While this boundary-agnostic segmentation can result in subword units being split at the edges of a segment, the CTC objective remains mathematically sound as it optimizes for token-level sequences rather than word-level boundaries. We repeated this training and alignment cycle for two iterations. 

    \item[Training Configuration] Pre-training and fine-tuning share a similar setup with slight differences.
        \begin{description}
            \item[Batch size and steps] 128 global batch size on 16 TPU v5e chips for both pre-training and fine-tuning. 1,000,000 pre-training steps and 300,000 fine-tuning steps.
            \item[Optimizer] Pre-training: AdaFactor \cite{shazeer2018adafactor} with a Noam schedule (0.01 peak learning rate, 10,000 warmup steps); gradients clipped to 0.5 global norm. fine-tuning: Adam \cite{kingma2014adam} with 0.001 learning rate.
            \item[Stability] 0.1 dropout in the Conformer encoder during both pre-training and fine-tuning. Exponential Moving Average with a 0.9999 decay rate during fine-tuning.
        \end{description}
\end{description}

\subsection{Pseudo-Streaming Inference for Long-Form Stability}
\label{sec:pseudo-streaming}

While models like Whisper \cite{radford2023robust} have advanced ASR significantly, they often exhibit instability when processing long-form audio. As identified by Bain et al. \cite{bain2023whisperx}, a primary failure mode in long-form ASR is \emph{drift}: a cumulative misalignment where the model's internal time-tracking or attention mechanisms begin to deviate from the actual acoustic signal. This often manifests as ``hallucination loops'', where the model repeats phrases, or ``deletion errors'', where large segments of audio are skipped entirely. In a medical context, such drift is catastrophic, as it can lead to the omission of critical clinical findings or the insertion of incorrect medication dosages. To mitigate these stability issues, we introduce \emph{Temporal Posterior Fusion}, a pseudo-streaming sliding-window inference algorithm.

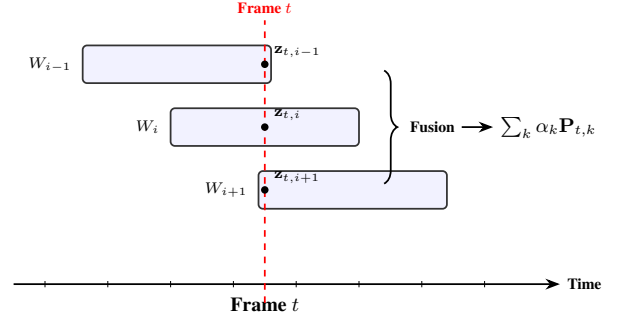
\begin{figure}[t]
\centering
\resizebox{\columnwidth}{!}{
\begin{tikzpicture}[
    window/.style={rectangle, draw=black!80, fill=blue!5, thick, minimum height=0.6cm, minimum width=3.0cm, rounded corners=2pt},
    dot/.style={circle, fill=black, inner sep=1.2pt},
    axis_label/.style={font=\scriptsize\itshape},
    math_node/.style={font=\scriptsize}
]

    \draw[thick, -{Stealth[scale=1.0]}] (-0.5,0) -- (8.2,0) node[right, font=\scriptsize\bfseries] {Time};
    
    \foreach \x in {0,1,2,3,4,5,6,7} {
        \draw (\x, 0.05) -- (\x, -0.05);
    }
    \node[below=0.1cm] at (3.5,0) (t_label) {\textbf{Frame $t$}};

    \def\yA{3.5} \def\yB{2.5} \def\yC{1.5}
    \def\centerline{3.5} 
    
    \node[window] (Wprev) at (\centerline-1.4, \yA) {};
    \node[left=0.05cm, font=\scriptsize] at (Wprev.west) {$W_{i-1}$};

    \node[window] (Wcurr) at (\centerline, \yB) {};
    \node[left=0.05cm, font=\scriptsize] at (Wcurr.west) {$W_i$};

    \node[window] (Wnext) at (\centerline+1.4, \yC) {};
    \node[left=0.05cm, font=\scriptsize] at (Wnext.west) {$W_{i+1}$};

    \draw[red, thick, dashed] (\centerline, -0.3) -- (\centerline, 4.2) node[above, font=\scriptsize\bfseries, text=red] {Frame $t$};

    \node[dot] (dot1) at (\centerline, \yA) {};
    \node[dot] (dot2) at (\centerline, \yB) {};
    \node[dot] (dot3) at (\centerline, \yC) {};
    
    \node[anchor=south west, math_node, xshift=1pt] at (dot1) {$\mathbf{z}_{t, i-1}$};
    \node[anchor=south west, math_node, xshift=1pt] at (dot2) {$\mathbf{z}_{t, i}$};
    \node[anchor=south west, math_node, xshift=1pt] at (dot3) {$\mathbf{z}_{t, i+1}$};

    \coordinate (bracket_tip) at (\centerline+2.0, \yB);
    
    \draw [decorate, decoration={brace, amplitude=6pt}, thick]
        (\centerline+1.9, \yA-0.1) -- (\centerline+1.9, \yC+0.1);

    \node[anchor=west, font=\scriptsize\bfseries] (fusion_txt) at (\centerline+2.2, \yB) {Fusion};
    
    \draw[-Stealth, thick] (fusion_txt.east) -- ++(0.5,0) node[right, font=\normalsize] {$\sum_k \alpha_k \mathbf{P}_{t,k}$};

    

\end{tikzpicture}
}
\caption{Temporal Fusion mechanism. Posterior logits $\mathbf{z}_{t,k}$ from different windows are fused via weighted averaging.}
\label{fig:fusion_flow}
\end{figure}

During long-form inference, an audio sequence $x$ is processed using a sliding window of fixed length $W$ and stride $S \leq W$.
We define the $k$-th window by its boundaries $[B_k, E_k)$, where the start of the window is $B_k = (k-1)S$, and the end is $E_k = B_k + W$.
For a frame at time $t$, $K_t = \{ k | B_k \leq t < E_k \}$ is the set of windows covering $t$.
As illustrated in Figure~\ref{fig:fusion_flow}, frame $t$ resides at a different relative position $t - B_k$ within each window $k \in K_t$.
This provides the model with diverse acoustic perspectives for the same frame: window $i-1$ provides rich left-context ($t$ being near its end), while window $i+1$ provides rich right-context ($t$ being near its start) .
 
To derive a unified frame-level posterior,
we aggregate these diverse perspectives using a weighted average. Let $\mathbf{P}_{t,k} \in \mathbb{R}^V$ be the posterior probability distribution generated by the $k$-th window for frame $t$, defined as:
\[
\mathbf{P}_{t,k} = \text{Softmax}(\mathbf{z}_{t,k})
\]
We define a weight vector $\mathbf{w} \in \mathbb{R}_{\geq 0}^W$ representing the importance of each relative position at which a frame might reside.
Having observed the first $a$ windows, the fused posterior $\mathbf{P}(y_t | x_{\leq a})$ is the weighted average of the individual posteriors of windows covering $t$ so far:
\[
\mathbf{P}_{\theta, a}(z_t | x) = \sum_{k=\min(K_T)}^{\min(K_T \cup \{a\})} \alpha_{t,k} \mathbf{P}_{t,k}
\]
where $\alpha_{t,k}$ is the normalized weight for the $k$-th window covering frame $t$:
\[
\alpha_{t,k} = \frac{w_{rel(t,k)}}{\sum_{k' \in K_T} w_{rel(t,k)}}
\]
where $rel(t,k) = t - B_k$ is the relative index of frame $t$ within window $k$.
We obtain the first $\mathbf{P}(y_t | x_{\leq a})$ when $a = \min(K_T)$, and $\mathbf{P}(y_t | x_{\leq a})$ ``converges'' after $a$ reaches $\max(K_T)$.

Given MedASR's compact 105M-parameter architecture, a large $S$ can be used for cheap offline inference. Alternatively, we can also afford a small high-frequency window stride $S$, to perform inference with low model latency while maintaining a large context within each window.

Through experimentation (Section~\ref{sec:hann-vs-uniform}), we found using a Hann window as $\mathbf{w}$ led to good results for different strides.

\section{Experiments}

\subsection{Test Sets}
\label{sec:test-sets}

We evaluate the performance of MedASR using both a publicly available test set (EyeGaze \cite{PhysioNet-egd-cxr-1.0.0, goldberger2000physiobank}), and 
held-out sets from our proprietary data (Section~\ref{sec:data}).

The proprietary evaluation sets are carefully curated to be speaker-independent, featuring approximately 5\% of the total unique speakers in the corpus with zero overlap between the training and testing partitions. This protocol ensures that the reported metrics reflect the model's ability to handle unseen vocal characteristics, accents, and acoustic environments typical of diverse clinical settings.

To avoid penalizing other models for not strictly following our textual format, we perform aggressive text normalization prior to WER scoring:
\begin{itemize}
    \item The following are removed: tags for de-identified spans; casing; puncuation; filler (``uh'', ``oh'') and ``unintelligble'' words;  voice commands that may be executed by some models (e.g. spoken puncutation; ``newline''; ``new paragraph'').
    \item Units are always abbreviated (e.g. ``millimeters'' becomes ``mm'').
    \item Single digit numbers are always in written form (e.g. ``two'' becomes ``2'').
\end{itemize}


\subsection{MedASR as an Offline Recognizer}
\label{sec:inference}

We compare MedASR (105M parameters) against two state-of-the-art foundational models: OpenAI Whisper (Large-v3) and Google Gemini 2.5 Pro.
For MedASR, we obtain the fused CTC lattices from sliding windows of length 20s \& stride 18s.
Each sliding window produces 500 logit vectors, which we fused using a size 500 Hann window (after discarding the boundary zeros) as weights $\mathbf{w}$.
We report results of both greedy decoding (no LM), and beam search with a 6-gram SentencePiece LM.
As shown in Table~\ref{tab:wer_results}, MedASR achieves a significant performance advantage, yielding a 58\% WER reduction on Eye Gaze compared to Whisper (12\% relative vs Gemini 2.5 Pro).

\begin{table*}
\centering
\caption{Comparison of WER across medical specialties
}
\label{tab:wer_results}
\begin{tabular}{lccccc}
\toprule
\textbf{Model} & \textbf{EyeGaze} & \textbf{RAD} & \textbf{FM} & \textbf{IM} & \textbf{GENINT}\\
\midrule
Whisper (Large-v3) & 12.5\% & 25.3\% & 32.5\% & 44.5\% & 33.1\% \\
Gemini 2.5 Pro     & 5.9\% & 10.0\% & 14.6\% & 21.3\% & 16.4\% \\
\midrule
\textbf{MedASR (no LM)} & 6.0\% & 5.4\% & 6.7\% & 9.9\% & 8.0\% \\
\textbf{MedASR (6-gram LM)} & \textbf{5.2\%} & \textbf{4.6\%} & \textbf{5.8\%} & \textbf{8.7\%} & \textbf{6.9\%} \\
\bottomrule
\end{tabular}
\end{table*}

The performance gap highlights the limitations of general-purpose models in specialized domains. While Whisper and Gemini exhibit high accuracy on standard benchmarks, they often struggle with the dense nomenclature and rapid pacing of clinical speech. MedASR’s stable performance across all four specialties suggests that its pre-training regime has successfully captured the underlying phonetic and semantic structures of medical discourse.

\begin{figure}
    \centering
    \caption{Offline MedASR (no LM) WER over different strides}
    \includegraphics[width=0.8\linewidth]{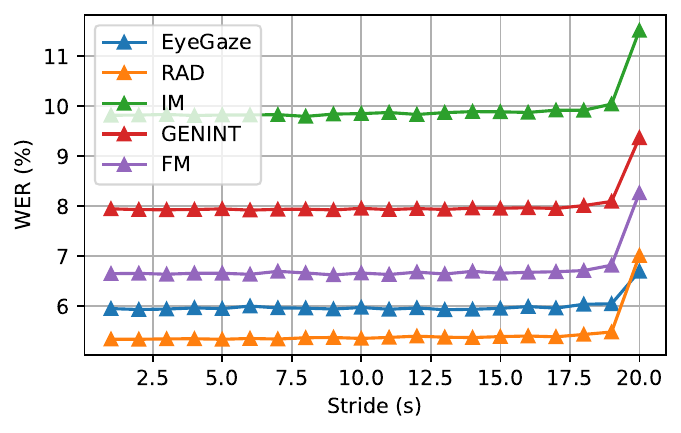}
    \label{fig:large-stride}
\end{figure}

A primary objective of our architecture is to resolve the ``drift'' issues identified in prior long-form ASR research \cite{bain2023whisperx}.
Holding the window size fixed at 20s, we investigate the sensitivity of the Word Error Rate of MedASR (no LM) to relatively large strides in Figure~\ref{fig:large-stride}.
MedASR's CTC architecture exhibited remarkable stability with respect to stride size, providing a robust guard against ``drift'' issues.

\subsection{Effective of Fusion Weights}
\label{sec:hann-vs-uniform}

\begin{figure}
    \centering
    \caption{Offline MedASR (no LM) WER using different fusion weights}
    \includegraphics[width=0.8\linewidth]{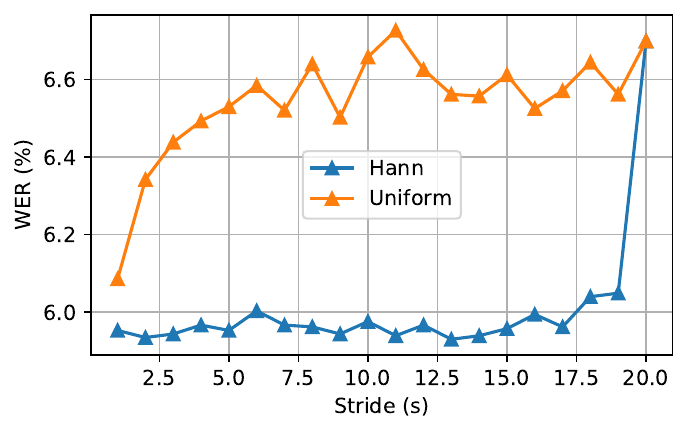}
    \label{fig:hann-vs-uniform}
\end{figure}

Figure~\ref{fig:hann-vs-uniform} compares WER on Eye Gaze of MedASR (no LM) fused using the Hann window and uniform weighting.
The Hann window exhibited superior WER consistently across different strides, ostensibly because it puts less weight when a frame has little left or right context.

\subsection{MedASR as a Streaming Recognizer}

For interactive use cases requiring low latency, MedASR can be used as a streaming recognizer with two simple changes to inference:
\begin{enumerate}
    \item Choose a small stride $S$ (e.g. 320ms) based on the latency and compute budget;
    \item Pad the start of the audio with $W$ seconds of zero-valued samples, so that the first sliding window ends at the very start of the audio (rather than time $W$).
\end{enumerate}

\begin{figure}
    \centering
    \caption{Streaming MedASR (no LM) WER over stride sizes}
    \includegraphics[width=0.8\linewidth]{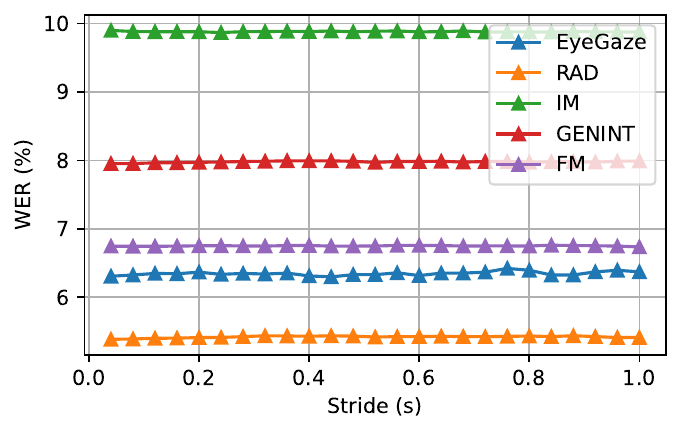}
    \label{fig:small-stride}
\end{figure}

Figure~\ref{fig:small-stride} shows this approach exhibits no significant increase in WER in most test sets\footnote{Except Eye Gaze, which saw a 0.3\% absolute increase in WER apparently due to the padding at start.} for MedASR (no LM), demonstrating that MedASR can be used with streaming inference without significantly decreasing WER.

\section{Conclusion}
We presented MedASR, an open-source ASR model optimized for long-form medical dictation. By utilizing Temporal Posterior Fusion within a pseudo-streaming framework, we successfully eliminated the ``drift'' and hallucination issues common in large-scale general-purpose models. Our experiments show a 58\% relative WER reduction over Whisper Large-v3, and confirm that the model maintains its accuracy with low model latency. This architecture provides a stable, high-precision solution for real-time clinical documentation without the computational overhead of offline foundational models.

\bibliographystyle{IEEEtran}
\bibliography{mybib}

\end{document}